Introducing Anisotropic Minkowski Functionals for Local Structure Analysis and

Prediction of Biomechanical Strength of Proximal Femur Specimens

By

Titas De

Submitted in Partial Fulfillment of the

Requirements for the Degree

Master of Science

Supervised by

Professor Axel W.E. Wismüller

Department of Electrical and Computer Engineering

Arts, Sciences and Engineering

Edmund A. Hajim School of Engineering and Applied Sciences

University of Rochester

Rochester, New York

2013

## Biographical Sketch

The author was born in Kolkata, India. He attended the Institute of Engineering and Management under West Bengal University of Technology, and graduated with a Bachelor of Technology degree in Electronics and Communication Engineering with an emphasis on Digital Signal and Image Processing in July 2011. He began interdisciplinary graduate studies in Electrical and Computer Engineering at the University of Rochester in August 2011 with continued study on Digital Signal Processing, Digital Image Processing, Computer Vision and Medical Imaging. He was awarded merit-based tuition scholarships from 2011 to 2013. During this time, he pursued research in Computational Radiology Lab under Dr. Wismüller (M.D., PhD), who himself is a radiologist.

The following presentations and publications were a result of work conducted during this study:

Axel Wismüller, Titas De, Eva Lochmüller, Felix Eckstein and Mahesh B. Nagarajan, "Introducing Anisotropic Minkowski Functionals and Quantitative Anisotropy Measures for Local Structure Analysis in Biomedical Imaging", Proceedings of SPIE Medical Imaging Conference, 2013.

## Acknowledgements

I would like to thank my committee members, Dr. Mark Bocko and Dr. Kevin Parker, and my advisor Dr. Axel Wismüller, for their attention and guidance during the course of my research.

Further, I greatly appreciate the direction and guidance received from my senior lab members Mahesh Nagarajan and Chien-chun Yang, from the Department of Biomedical Engineering. Their guidance and direction helped me successfully overcome the challenges and the difficulties encountered throughout my thesis research and academic life in Rochester, NY.

Lastly, I would like to thank all my collaborators, i.e. professors, faculty members and researchers from different parts of the world for their help and support. They include Dr. Felix Eckstein and Dr. Eva Lochmüller from Paracelsus Medical University at Salzburg, Austria.

Finally, I would also like to thank the Department of Electrical and Computer Engineering at the University of Rochester for their support through tuition scholarships.

## Abstract

Bone fragility and fracture caused by osteoporosis or injury are prevalent in adults over the age of 50 and can reduce their quality of life. Hence, predicting the biomechanical bone strength, specifically of the proximal femur, through non-invasive imaging-based methods is an important goal for the diagnosis of Osteoporosis as well as estimating fracture risk. Dual X-ray absorptiometry (DXA) has been used as a standard clinical procedure for assessment and diagnosis of bone strength and osteoporosis through bone mineral density (BMD) measurements. However, previous studies have shown that quantitative computer tomography (QCT) can be more sensitive and specific to trabecular bone characterization because it reduces the overlap effects and interferences from the surrounding soft tissue and cortical shell.

This study proposes a new method to predict the bone strength of proximal femur specimens from quantitative multi-detector computer tomography (MDCT) images. Texture analysis methods such as conventional statistical moments (BMD mean), Isotropic Minkowski Functionals (IMF) and Anisotropic Minkowski Functionals (AMF) are used to quantify BMD properties of the trabecular bone micro-architecture. Combinations of these extracted features are then used to predict the biomechanical strength of the femur specimens using sophisticated machine learning techniques such as multi-regression (MultiReg) and support vector regression with linear kernel (SVRlin). The prediction performance achieved with these feature sets is compared to the standard approach that uses the mean BMD of the specimens and multi-regression models using root mean square error (RMSE).

The best prediction performance using Anisotropic Minkowski Functionals (AMF) gives RMSE = 0.904 ± 0.105, which is significantly better than the ones obtained using Isotropic Minkowski Functionals (RMSE = 1.585 ± 0.167) and DXA BMD (RMSE = 0.960 ± 0.131).

Contributors and Funding Sources

This research was funded in part by the Clinical and Translational Science Award 5-28527 within the Upstate New York Translational Research Network (UNYTRN) of the Clinical and Translational Science Institute (CTSI), University of Rochester, and by the Center for Emerging and Innovative Sciences (CEIS), a NYSTAR-designated center for Advanced Technology.

## Table of Contents

## List of Tables

List of Figures

**Chapter 1**

**Introduction**

**1.1    Motivation for this work**

Examining and interpreting medical images such as MRI and CT can be a tedious and exhaustive task for radiologists; extraction of relevant and precise clinical findings for correct clinical decision-making requires tremendous training and clinical experience [1]. In spite of their training and experience, clinical findings can be overlooked or misinterpreted for various reasons including distraction, reader fatigue, anatomical structure overlapping, etc. [2-6]. In addition, the interpretation is also subject to inter-observer variations which can lead to incorrect decisions as well. Finally, the native constraint of human eye-brain visual system also places certain limitations on the ability of radiologists in discerning and recognizing brightness, morphology and patterns on medical images [7, 8]. As a consequence, it imposes challenges and difficulties in making precise and objective interpretation while evaluating clinical findings.

Computer-aided diagnosis system (CADx), on the other hand, allows extraction and analysis of image features that are inaccessible to human eye-brain visual system and thus, provide a more objective and consistent decision, which can be used as a complementary opinion to help radiologists in clinical evaluations [4,8]. This proposal aims to propose a novel CADx system in the skeleton disease setting, in order to improve accuracy in the diagnosis of osteoporosis and fracture risk prediction. A brief summary of CADx mechanism, research background of the relevant diagnostic modality, and the proposed solution are described below.

## 1.2    Computer-Aided Diagnosis Principles and Mechanisms

CADx [1-6], as used in this study, can be divided into four stages: region/volume of interest (ROI/VOI) selection, texture feature extraction, decision/regression determining algorithm, and decision output, all of which are described below.

CADx usually begins with ROI/VOI selection – regions that contain relevant information for clinical findings such as lesions, or anatomical sites such as vertebral body or femur head are selected for further detailed investigation. ROI/VOI selection can be accomplished by manual, semi-automatic or fully automatic methods.

Texture feature extraction utilizes texture feature analysis methods, evolved over many years, to extract quantified features that characterize patterns on an image. Some popular texture analysis methods include conventional statistics, Minkowski Functionals (MF), Gray-Level Co-occurrence Matrix (GLCM) [9, 10] and Scaling Index Method (SIM) [11-13]. Although many texture feature analysis methods exist, the ultimate purpose is similar – to extract features from the ROIs/VOIs of medical images.

The extracted features are used to construct a mathematical model using a decision/regression algorithm, also known as *machine learning* algorithm. This model is subsequently used to provide quantitative analysis for undetermined cases where the designated features are provided.

The outcome of such a system can serve as a complementary opinion and assist radiologists in clinical decision making.

## 1.3    Research Background and Focus

<u>Research Background</u>

Osteoporosis, a disease related to the imbalance between trabecular bone formation and resorption, is one of the most common age related diseases targeting elderly people [18]. The progression of osteoporosis can lead to osteoporotic fractures, which not only reduces the quality of life but also increases the mortality rate [18]. Previous studies have predicted that the osteoporotic fracture risk population will reach 6.26 million worldwide by the year 2050 [19, 20]. Thus, accurate prediction of osteoporotic fracture risks is an important aid for clinical assessment and management of osteoporosis [21-25].

Dual-energy X-ray absorptiometry (DXA) has been the standard technique for measuring bone quality in terms of bone mineral density (BMD) for purposes of osteoporotic fracture risk estimation [20-24]. BMD measurements through DXA at the site of the proximal femur have shown to be highly predictive of bone fractures when compared to other sites [24-27]. Such BMD measurements can contribute to increased accuracy in bone fracture risk assessment at the hip.

However, BMD measurements alone do not account for a complete profile of the trabecular bone microarchitecture; thus leading to some inconsistency in osteoporosis diagnosis. Kanis et al. suggested that the presence of normal values of BMD within the average range does not necessarily indicate the absence of osteoporosis but rather a lower risk of developing osteoporosis or related fractures [25, 29].  In fact, BMD measurements for people with and without prevalent femur fractures have been shown to overlap, which indicates that other factors need to be taken into account for bone strength estimation [29]. In addition, previous studies have also suggested that DXA-derived BMD measurements are adversely affected by interference from surrounding cortical shell, adipose tissue and soft tissue, which result in inaccuracies for bone strength estimation and mislead the diagnostic interpretation [26-28, 21-35].

Quantitative computer tomography (QCT), in contrast with DXA, can be used to eliminate any interference from surrounding tissue and allow a direct and independent estimation of either the cortical or trabecular compartment; thus providing an exclusive measure of BMD in the trabecular compartment. Therefore, QCT can be used to successfully improve the efficacy of bone loss and fracture risk assessment, which has been previously demonstrated in the spinal fracture studies. In fact, such studies showed that QCT measurements in the central trabecular region of interest excluded sources of error such as osteophytes and hypertrophic posterior elements, which may artificially elevate integral BMD measures and reduce diagnostic efficacy [34-37].

Research Focus

Although BMD measured by QCT is strongly correlated with fracture risk, it is still not a satisfactory predictor for bone strength due to variations in bone morphology and structure [38]. Therefore, improving the accuracy of in-vivo estimation of the biomechanical strength of proximal femurs through novel techniques is an important goal in osteoporosis research. In this regard, previous studies have reported that QCT-derived BMD, when used in combination with anatomical variables such as bone volume, trabecular separation or femoral hip axis length (HAL), exhibit better bone strength estimation over the DXA-derived BMD in the femur [37-39]. Such findings indicate that bone features other than BMD may also play a role in determining bone strength [32, 37-39].

Therefore, we propose an improved characterization of trabecular bone, as visualized on multi-detector CT images, with higher order geometric feature vectors derived from Isotropic and Anisotropic Minkowski Functionals [9, 10]. Such features, along with conventionally used BMD measurements, are then used to construct bone strength prediction models with different supervised machine learning techniques such as multi-regression and support vector regression with linear kernel [13-16], and the ability of such models to predict the bone strength is evaluated. The following figure gives an overview summary of the experiment setup and data presented in this research.

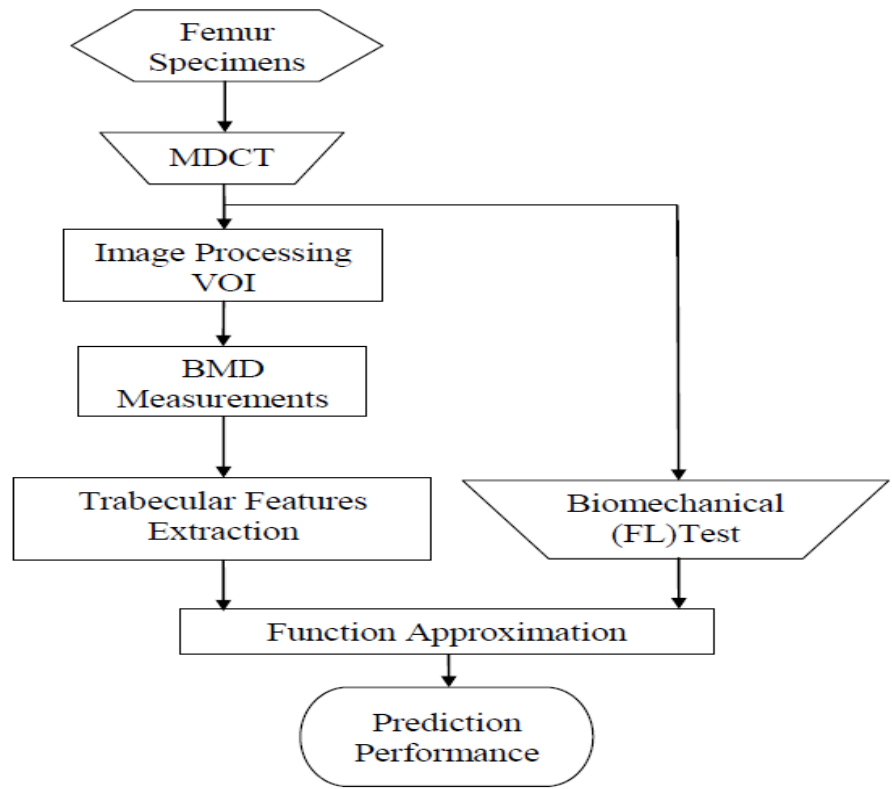

**Figure 1.1 -** Overview of my experimental setup and methods used. The trabecular features (BMD mean, Isotropic and Anisotropic Minkowski Functionals) were extracted from VOIs annotated on MDCT images of the femur specimens post-processed to facilitate conversion of intensity values from Hounsfield units to BMD values. Two function approximation methods, i.e. multi-regression and support vector regression analysis, were then used to predict the failure load (FL); the similarity between predicted FL and actual values determined through biomechanical testing was quantified through RMSE.

## 1.4    Experimental Materials and Data

This section describes our experimental materials and the relevant pre-processing procedures. These include femur specimens, imaging modalities, VOI selection and biomechanical test, and bone mineral density unit conversion.

<u>Femur Specimens</u>

Left femoral specimens were harvested from fixed human cadavers over a time period of four years. The donors had dedicated their body to the investigators at the Institute of Anatomy and Musculoskeletal Research, Paracelsus Private Medical University Salzburg for educational and research purposes prior to death, in line with local institutional and legislative requirements. To exclude donors with diffuse metastatic bone disease and hematological or metabolic bone disorders other than osteoporosis, biopsies were obtained from the iliac crest and examined histologically as part of the general research protocol. The histological assessment was performed by a surgeon who had been trained as a pathologist for 3 years with a focus on bone pathology. Specimens where signs of fractures were detected either in radiographs or during preparation as well as specimens that displayed a fracture of the femoral shaft (rather than of the proximal femur) during the mechanical test were excluded. Using the above criteria, a subset of 146 human femur specimens were used for this study. The bones were removed from the cadavers with a variable amount of surrounding soft tissues. To create uniform scanning conditions, the soft tissue surrounding the bones was removed for imaging and biomechanical testing.

## Multi-Detector Computed Tomography (MDCT)

Cross-sectional images of the femora were acquired using a 16-row multi-detector (MD)-CT scanner (Sensation 16; Siemens Medical Solutions, Erlangen, Germany). The specimens were placed in plastic bags filled with 4% formalin/water solution. Air was removed with a vacuum pump and plastic bags were sealed. These were positioned in the scanner as in an in-vivo exam of the pelvis and proximal femur with mild internal rotation of the femur. Each specimen was scanned once, except for 3 specimens who were scanned twice for precision measurements, with a protocol using collimation and table feed of 0.75 mm, and a reconstruction index of 0.5 mm. A high resolution reconstruction algorithm (kernel U70u) was used, resulting in an in-plane resolution of 0.19 x 0.19 mm$^2$ and anisotropic voxel size of 0.19 x 0.19 x 0.5 mm$^3$. A kilovolt peak of 120 kVp was used with 100 mA. The image matrix was 512 x 512 pixels, with a field of view of 100 mm. For calibration purposes, a reference phantom (Osteo Phantom, Siemens) was placed below the specimens (Fig. 1.2)

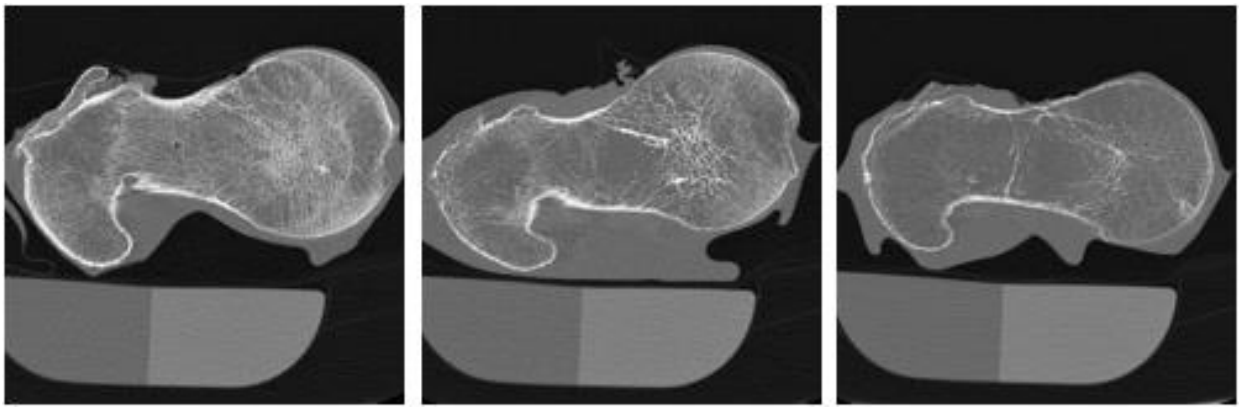

**Figure 1.2 -** MDCT images of selected femur specimens. From left to right, the specimens are categorized as high, medium and low biomechanical strength, respectively based on failure load tests. The osteo phantom used for each specimen is also shown at the bottom.

Image Processing and Volume of Interest (VOI) Selection

The outer surface of the cortical shell of the femur was segmented by using bone attenuations of the phantom in each image. The specimens were segmented automatically; however, the shape of the binary mask was manually corrected if errors in segmentation occurred due to a thin cortical shell caused by high grade focal bone loss or to adjacent anatomic structures such as blood vessels penetrating the cortex. The corrections for all specimens were performed by one of two radiologists. Based on a priori knowledge about the orientation of the specimens in the CT scans, the superior part of the femoral head was identified automatically. Based on the size and shape of the contours and the center of mass of the contours of consecutive slices, the superior part of the femoral head was detected. A sphere was fitted to the superior surface points of the femoral head using a Gaussian Newton Least Squares technique. The fitted sphere was scaled down to 75% of its original size to account for cortical bone and shape irregularities like the fovea capitis, and then saved as the femoral head volume of interest (VOI). Because a cylinder can approximate the shape of the femur neck, with a similar procedure of head VOI selection, a cylindrical VOI was computed and automatically fitted to the neck region. The resulting cylinder was saved as the femur neck VOI.

For the trochanter VOI selection, a cone-like shape VOI was fitted into the trochanter region based on the bone surface points relative to the neck axis, the surface regions corresponding to the trochanter, inferior part of the neck and superior part of the shaft. Main eigenvector of these regions was used as an initial estimate of the axis of a cone that was fitted to the bone surface points in these regions. Bone surface points in these regions were matched to the fitted cone axis and to the original neck axis. The trochanter bone surface points were then saved as the trochanter VOI. Further detail of the VOI selection algorithms can be referred to Huber et al. [43].

Biomechanical Tests

The failure load was assessed using a side-impact test, simulating a lateral fall on the greater trochanter as described in paper [54]. Briefly, the femoral shaft and head faced downward could be moved independently of one another while the load was applied on the greater trochanter using a universal materials testing machine (Zwick 1445, Ulm, Germany) with a 10kN force sensor and dedicated software. The failure load was defined as the peak of the load-deformation curve.

VOI extraction and BMD conversion

The first step was to extract the trabecular VOI from original MDCT images (shown in Fig. 2). These MDCT images were segmented by the pre-defined VOIs with respect to the head, neck and trochanter regions. Three different shapes of VOIs (sphere, cylinder and cone) were designed to fit into different regions (head, neck and trochanter) of the femur specimens (Fig. 3) as described in Huber et al. [43]. Within each of the extracted VOIs the Hounsfield Unit (HU) is converted into BMD unit ($mg/cm^3$) based on the HU value of the Osteo calibration phantom and the following equation:

$$BMD = [HA_B/ (HU_B - HU_W)] * (HU - HU_W), \ \ldots\ldots \ \ (1)$$

Each of the above variables is explained below.

The calibration phantom is composed of two portions of hydroxyapatite which contains the hydroxyapatite density values of $HA_W$ = 0 $mg/cm^3$ and $HA_B$ = 200 $mg/cm^3$ for the water-like and bone-like parts of the calibration phantom, respectively. In addition to these constants, $HU_W$ and $HU_B$ are the attenuations (HU readings) from the MDCT image for water-like and bone-like parts of the phantom, respectively. So, the HU values of the water-like and bone-like phantom were recorded for each slice throughout the scan.

The following table provides a brief table of HU readings for selected substances:

| Substance | HU |
| --- | --- |
| Air | -1000 |
| Fat | -84 |
| Water | 0 |
| Blood | +35 to +45 |
| Muscle | +40 |
| Soft Tissue | 100 to 300 |
| Bone | +700(cancellous bone) to +3000(dense bone) |

**Table 1.1 -** Hounsfield Unit readings for selected substances. Air tend to have large negative HU readings; whereas, fat has minor negative HU reading. Soft tissue has HU reading between 100 and 300. Bone tissue, depends on the density, and has HU reading from 700 to 3000. Note that the BMD of trabecular region has range between [-300 1400] after converting from HU readings to BMD.

After segmentation, the Hounsfield Unit images within the VOIs were converted into the BMD unit images (Figure 1.3).

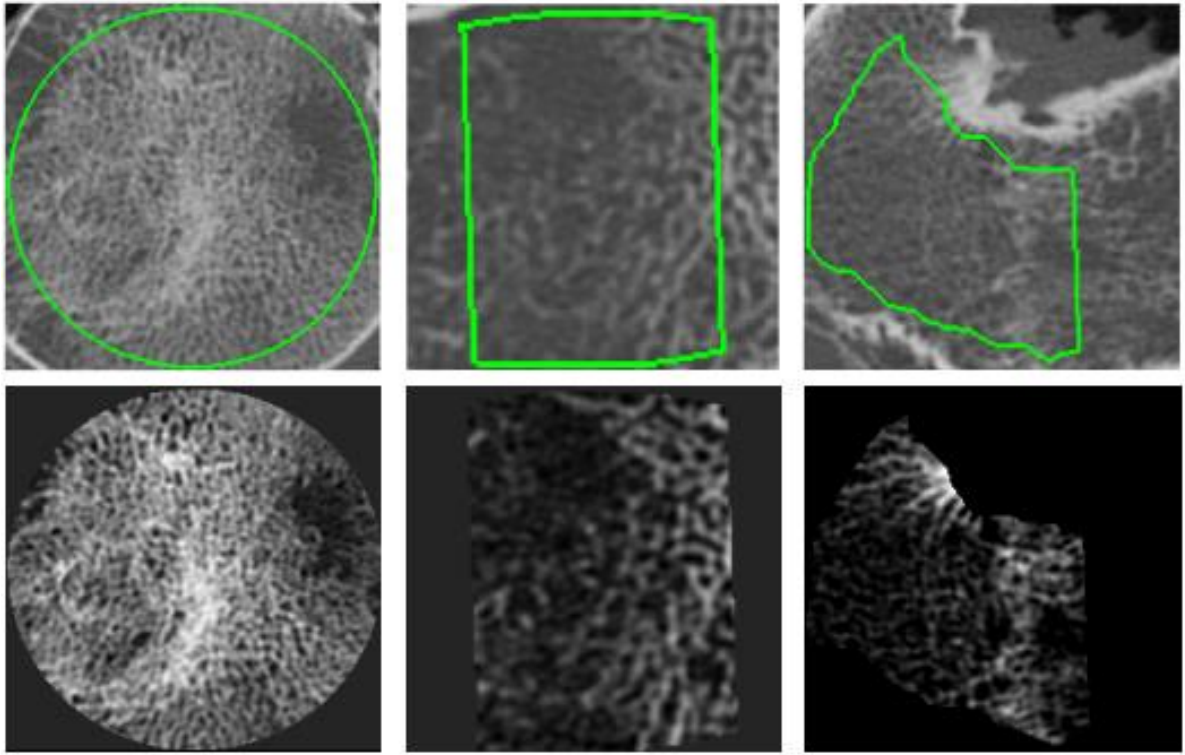

**Figure 1.3 -** Results of ROI-fitting and BMD conversion in selected specimens. ROI fitting and BMD conversion in specimens shown in Figure 1. (Top row) Three shapes (circle, quadrilateral and irregular shape) of ROIs were fit into the head, neck and trochanter region of femur specimens, respectively. ROI boundaries are overlaid on the corresponding MDCT images of the three regions. From left to right are head, neck and trochanter. Note the three images are not shown in the consistent scales since head region is the largest, trochanter second and neck being the smallest. (Bottom row) Hounsfield Unit (HU) images within each ROI are converted to corresponding BMD values.

After the ROI selection and the BMD conversion, the BMD images are then ready for feature extraction and analysis, which are discussed in Chapter 2.

**Chapter 2**

**Feature Analysis**

Feature analysis techniques are utilized to represent massive original or raw information, as found in medical imaging (for example), in a more compact and concise manner. As soon as one can represent the large volume of medical images with a compact size of features, these features can then be used to construct mathematical models with machine learning techniques.

This chapter describes three different feature extraction techniques used to characterize the femur BMD images in this study. These methods include the conventional statistical features, Isotropic Minkowski Functionals (IMF) and Anisotropic Minkowski Functionals (AMF).

## 2.1   Conventional Statistical Features

Conventional statistical features are usually the most common and the simplest features used in pattern recognition. Here, the BMD distributions within VOIs for 3D images and ROIs for 2D images are characterized by their statistical moments. We have Dual Energy X-ray Absorptiometry (DXA) Bone Mineral Density (BMD) images which are 2D images, as opposed to Quantitative Computed Tomography (QCT) Bone Mineral Density (BMD) images which are actually 3D images. The current clinical standard for bone density evaluation is using DXA BMD obtained from 2D DXA image of the bone. But we will be extracting all kinds of morphometric features (IMF and AMF) from 3D QCT images.

## 2.2   Minkowski Functionals

The concept of Minkowski Functionals is explained in detail in the paper "INTEGRAL-GEOMETRY MORPHOLOGICAL IMAGE ANALYSIS." (Michielsen, De Raedt) [9]. In short, if we have a 2D image black and white image, we can find out the 3 Minkowski Functionals (Area, Perimeter and Euler characteristic) from that whole black and white image using the following formula :

Area = $n_s$,  Perimeter = $-4n_s + 2n_e$, Euler Characteristic = $n_s - n_e + n_v$

Here $n_s$ = the total number of white pixels, $n_e$ = total number of edges, and $n_v$ = total number of vertices.

Similarly, if we have a 3D black and white image, we can find out the 4 Minkowski Functionals (Volume, Surface Area or Surface, Mean Breadth and Euler Characteristic) from the entire black and white 3D image volume by using the following formula:

Volume = $n_s$ , Surface = $-6n_s + 2n_f$ , Mean Breadth = $3n_s - 2n_f + n_{e,}$ , and Euler Characteristic = $-n_s + n_f - n_e + n_v$

Here $n_s$ = the total number of white pixels, $n_e$ = total number of edges, $n_v$ = total number of vertices and $n_f$ = total number of faces.

## 2.3    Isotropic Minkowski Functionals

We already know there are four Minkowski Functionals (MFs) for a 3D image which are Volume, Surface, Mean Breadth and Euler Characteristic which measures the topological characteristic of the entire image as a whole. But in my study, instead of calculating the Minkowski Functionals for the entire 3D images, I will calculate it for each white voxel in the binary image using the information about the local neighborhood of that voxel. The neighborhood voxels including the central voxel are first weighted by a pre-defined kernel of the same size as the neighborhood, and these resultant weighted voxels are used to calculate the kernel Minkowski Functionals (as may be called). Thus instead of getting just one value corresponding to each Minkowski Functional, we now get a vector (column) of values and the size of the vector depends on the number of white voxels in the image.

Let me give you an example. Say I have a 3D black and white (binary) image with white voxels represented by 1s and black voxels by 0s. Say the size of the image is M x N x P. Let's say the total number of white voxels in the image is $N_{WP}$ (< M*N*P).  Let's say we are using a kernel of dimensions m x n x p to compute the kernel Minkowski functionals. The output which we get would be a set of 4-D row vectors, with each vector containing the Volume, Surface, Mean Breadth and Euler Characteristic values for each voxel obtained using the above-mentioned kernel of size m x n x p. The number of such row vectors would be number of white voxels in the image, and which is $N_{WP}$. In short, our output would be a $N_{WP}$ x 4 matrix.

Choice of a suitable kernel is a very important task, as these are used to describe the local texture features in the image. The simplest kernel to use would be a plain cubic kernel with all weights equal to 1. If we use such a kernel, we notice that such a kernel is isotropic in nature i.e. it does not change its shape if we rotate it in any direction. We have named the kernel Minkowski Functionals obtained using an isotropic kernel (such as a plain cubic kernel) as isotropic Minkowski Functionals. Talking about isotropic kernels, we can also use a Gaussian kernel which can be made isotropic or rotation invariant by having its standard deviation in all the three axes (x, y and z) as the same.

## 2.4  Anisotropic Minkowski Functionals

We now wanted to impose anisotropy or specificity of direction in the measurement of our Minkowski Functionals. So instead of using a Gaussian kernel which is rotation invariant i.e. having the same standard deviation in all the three axes, we are using Gaussian kernels which have a longer standard deviation in a specific direction as compared to the two other orthogonal directions. (Note - The three directions does not have to be only x, y and z axes. They can be any three orthogonal directions in the 3-D space). As before, we are calculating the kernel Minkowski Functionals for each white voxel, but this time for a number of different direction-oriented kernels.

Before we discuss any further, let me talk about direction-oriented kernels. Even though, we are talking about 3D images and co-ordinates, but I will try to explain the direction orientation in 2D, as it will make things simple.

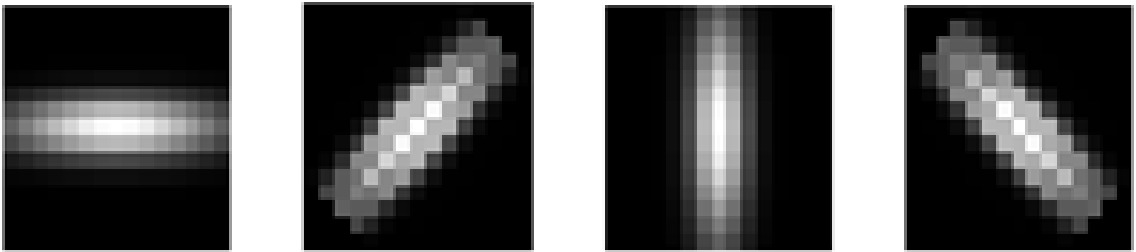

**Figure 2.1:** Figure showing 2D Gaussian kernels oriented in different directions.

The above picture shows 4 Gaussian kernels oriented at angles 0, 45, 90 and 135 degrees respectively. By looking at the above picture, you can have a sense of how orientation in 3D co-ordinate space would look like. The difference between 2D and 3D orientation is that in 2D you care about only one angle i.e theta (which is the angle between the projections in the x and y axes), while in 3D you care not only about theta, but also about phi (which is the angle between projections in the xy plane and the z axis).

So we are using Gaussian kernels oriented in different directions in 3D space to calculate the kernel-wise Minkowski Functionals. At the end, what we get corresponding to each Minkowski Functionals (Volume, Surface, Mean Breadth and Euler Characteristic) is a set of vectors (columns) containing the Minkowski Functional values for each direction. In short, each white voxel has a set of values for each Minkowski Functionals. We then use these set of values and Principal Component Analysis to find the resultant angles [ a) theta - angle between projected values in the x and y axes; b) phi - angle between projected values in the z axes and the xy plane] and also the fractional anisotropy (degree of anisotropy or direction specificity) for each white voxel. Fractional Anisotropy (FA) is obtained using the formula

$$FA = \frac{\sqrt{(\lambda_1 - \lambda_2)^2 + (\lambda_1 - \lambda_3)^2 + (\lambda_2 - \lambda_3)^2}}{\sqrt{2(\lambda_1^2 + \lambda_2^2 + \lambda_3^2)}}$$

Basically what it comes to is that for each Minkowski Functional, we now have 3 vectors (columns) which are theta, phi and the fractional anisotropy (FA). Now theta and phi can contain values only between 0 and 180 degrees, and FA can have values only between 0 and 1. This is in contrast to the isotropic Minkowski Functionals where the minimum and maximum limits are subject to the local structure of the 3D image and also the size and characteristic of the kernel used.

## 2.5   Features obtained for Prediction Performance

After getting the FA, theta and phi vectors for each minkowski functional, we are extracting histogram of values from them with pre-defined bin centers. These histograms are the Anisotropic Minkowski Functionals (AMF) features which are used in our research for prediction performance. For Isotropic Minkowski Functionals, which do not have universal minimum and maximum limits, we are first finding the min and max limits from the training set. Then we are using these limits to define the bin centers of the histogram. Ultimately we are obtaining the histogram features from the entire dataset which is then fed into machine learning techniques to obtain prediction performance.

**Chapter 3**

**Machine Learning Algorithms**

**3.1  Introduction**

Two definitions of Machine Learning are offered. Arthur Samuel described it as: "the field of study that gives computers the ability to learn without being explicitly programmed." This is an older, informal definition.

Tom Mitchell provides a more modern definition: "A computer program is said to learn from experience E with respect to some class of tasks T and performance measure P, if its performance at tasks in T, as measured by P, improves with experience E."

Example: playing checkers.

- E = the experience of playing many games of checkers

- T = the task of playing checkers.

- P = the probability that the program will win the next game.

Machine Learning involves programming computerized mathematical models to optimize a performance criterion using example training data or past experience. Such models are defined with weight parameters in the sense of weighting the importance of different attributes or features. The model may be predictive i.e. to make future predictions, or descriptive, i.e. to gain knowledge from data, or both [49].

Machine Learning uses the theory of statistics in building mathematical models, denoted as Solution, Decision Function, Target Function, Hypothesis or Classifiers, where the core task is drawing inferences from a sample. The role of computer science is two-fold: First, in training, we need efficient algorithms, known as learning algorithm, to solve the optimization problem, as well as to store and process the massive amount of training data or training set we generally have. Second, once a model is learned, its representation and algorithmic solution for inference or prediction needs to be efficient as well. In certain applications, the efficiency of the learning or inference algorithms, namely, its space and time complexity, may be as important as its predictive accuracy [50].

## Supervised Learning

In supervised learning, we are given a data set and already know what our correct output should look like, having the idea that there is a relationship between the input and the output [51].

Supervised learning problems are categorized into "regression" and "classification" problems. In a regression problem, we are trying to predict results within a continuous output, meaning that we are trying to map input variables to some continuous function. In a classification problem, we are instead trying to predict results in a discrete output. In other words, we are trying to map input variables into discrete categories.

Examples: given data about the size of houses on the real estate market, try to predict their price. Price as a function of size is a continuous output, so this is a regression problem.

We could turn this example into a classification problem by instead making our output about whether the house "sells for more or less than the asking price." Here we are classifying the houses based on price into two discrete categories.

Unsupervised Learning

Unsupervised learning, on the other hand, allows us to approach problems with little or no idea what our results should look like. We can derive structure from data where we don't necessarily know the effect of the variables [51].

We can derive this structure by clustering the data based on relationships among the variables in the data.

With unsupervised learning there is no feedback based on the prediction results, i.e., there is no teacher to correct you. It's not just about clustering. For example, associative memory is unsupervised learning.

Examples: Clustering.-Take a collection of 1000 essays written on the US Economy, and find a way to automatically group these essays into a small number that are somehow similar or related by different variables, such as word frequency, sentence length, page count, and so on.

Suppose a doctor over years of experience forms associations in his mind between patient characteristics and illnesses that they have. If a new patient shows up then based on this patient's characteristics such as symptoms, family medical history, physical attributes, mental outlook, etc. the doctor associates possible illness or illnesses based on what the doctor has seen before with similar patients. This is not the same as rule based reasoning as in expert systems. In this case we would like to estimate a mapping function from patient characteristics into illnesses.

### 3.2 Linear Regression with one variable

<u>Model Representation</u>

Recall that in regression problems, we are taking input variables and trying to map the output onto a continuous expected result function.

Linear regression with one variable is also known as "univariate linear regression." Univariate linear regression is used when you want to predict a single output value from a single input value. We're doing supervised learning here, so that means we already have an idea what the input/output cause and effect should be.

Our hypothesis function has the general form: $h_\theta(x) = \theta_0 + \theta_1 x$. We assign $h_\theta$ with values for $\theta_0$ and $\theta_1$ to get our output 'y'. In other words we are trying to create a function called $h_\theta$ that is able to reliably map our input data (the x's) to our output data (the y's).

Example:

| x (input) | y (output) |
|-----------|------------|
| 0 | 4 |
| 1 | 7 |
| 2 | 7 |
| 3 | 8 |

Now we can make a random guess about our $h_\theta$ function: $\theta_0 = 2$ and $\theta_1 = 2$.

The hypothesis function becomes $h_\theta(x) = 2 + 2x$.

So for input of 1 to our hypothesis, y will be 4. This is off by 3.

## Cost Function

We can measure the accuracy of our hypothesis function by using a cost function. This takes an average (actually a sophisticated version of an average) of all the results of the hypothesis with inputs from x's compared to the actual output y's.

The cost function is:

$J(\theta_0, \theta_1) = \frac{1}{2m} \sum_{i=1}^{m} (h_\theta(x^{(i)}) - y^{(i)})^2$, where m is the size of the training set.

You can think of this equation as taking the average of the differences of all the results of our hypothesis and the actual correct results.

Now we are able to concretely measure the accuracy of our predictor function against the correct results we have, so that we can predict new results we don't have.

## Gradient Descent

So we have our hypothesis function and we have a way of measuring how accurate it is. Now what we need is a way to automatically improve our hypothesis function. That's where gradient descent comes in.

Imagine that we graph our hypothesis function based on its parameters $\theta_0$ and $\theta_1$ (actually we are graphing the cost function for the combinations of parameters). This can be kind of confusing; we are moving up to a higher level of abstraction. We are not graphing x and y itself, but the guesses of our hypothesis function.

We put $\theta_0$ on the x axis and $\theta_1$ on the z axis, with the cost function on the vertical y axis. The points on our graph will be the result of the cost function using our hypothesis with those specific θ parameters.

We will know that we have succeeded when our cost function is at the very bottom of the pits in our graph and our result is 0 (or close to it).

The way we do this is by taking the derivative (the line tangent to a function) of our cost function. The slope of the tangent is the derivative at that point and it will give us a direction to move towards. We step down that derivative by a constant value called alpha (α).

The gradient descent equation is:

repeat until convergence:

$\theta_j := \theta_j - \alpha \frac{\partial}{\partial \theta_j} J(\theta_0, \theta_1)$   , for j = 0 and j = 1

Intuitively, this could be thought of as:

repeat until convergence:

$\theta_j := \theta_j - \alpha \left[ \text{slope or derivative} \right]$

Gradient Descent for Linear Regression

When specifically applied to the case of linear regression, a new form of the gradient descent equation can be derived. We can substitute our actual cost function and our actual hypothesis function and modify the equation to:

repeat until convergence: {

$\theta_0 := \theta_0 - \alpha \sum_{i=1}^{m} \left( h_\theta \left( x^{(i)} \right) - y^{(i)} \right)$

$\theta_1 := \theta_1 - \alpha \sum_{i=1}^{m} \left( \left( h_\theta \left( x^{(i)} \right) - y^{(i)} \right) x^{(i)} \right)$ }

Over here m is the size of the training set, $\theta_0$ is a constant that will be changing simultaneously with $\theta_1$ and $x^{(i)}, y^{(i)}$ are values of the given training set (data).

Note that we have separated out the two cases for $\theta_j$ and that for $\theta_1$ we are multiplying $x^{(i)}$ at the end due to the derivative.

The point of all this is that if we start with a guess for our hypothesis and then repeatedly apply these gradient descent equations, our hypothesis will become more and more accurate.

### 3.3  Linear Regression with Multiple variables (Multi-Regression)

Multiple Features

Linear regression with multiple variables is also known as "multivariate linear regression."

We now introduce notation for equations where we can have any number of input variables.

$x_j^{(i)}$ = value of feature j in the $i^{th}$ training example.

$x^{(i)}$ = the column vector of all the input features of the $i^{th}$ training example

m = the number of training examples

n = $|x^{(i)}|$ (the number of features)

Now define the multivariable form of the hypothesis function as follows, accommodating these multiple features:

$h_\theta(x) = \theta_0 + \theta_1 x_1 + \theta_2 x_2 + \theta_3 x_3 + \cdots + \theta_n x_n$

Using the definition of matrix multiplication, our multivariable hypothesis function can be concisely represented as:

$$h_\theta(x) = [\ \theta_0\ \theta_1 ... \theta_n\ ] \begin{bmatrix} x_0 \\ x_1 \\ \vdots \\ x_n \end{bmatrix} = \theta^T x$$

This is a vectorization of our hypothesis function for one training example.

Now we can collect all m training examples each with n features and record them in an n+1 by m matrix. In this matrix we let the values of the subscript (feature) represent the row number (except the initial row is the "zeroth" row), and the values of the superscript (the training example) represent the column number, as shown in the next page:

$$X = \begin{bmatrix} x_0^{(1)} & x_0^{(2)} & \cdots & x_0^{(m)} \\ x_1^{(1)} & x_1^{(2)} & \cdots & x_1^{(m)} \\ \vdots & \vdots & & \vdots \\ x_n^{(1)} & x_n^{(2)} & \cdots & x_n^{(m)} \end{bmatrix} = \begin{bmatrix} 1 & 1 & \cdots & 1 \\ x_1^{(1)} & x_1^{(2)} & \cdots & x_1^{(m)} \\ \vdots & \vdots & & \vdots \\ x_n^{(1)} & x_n^{(2)} & \cdots & x_n^{(m)} \end{bmatrix}$$

Notice above that the first column is the first training example, the second column is the second training example, and so forth.

Now we can define $h_\theta(x)$ as a row vector that gives the value of $h_\theta(x)$ at each of the m training examples:

$$h_\theta(x) = [\theta_0 x_0^{(1)} + \theta_1 x_1^{(1)} + \theta_2 x_2^{(1)} + .... + \theta_n x_n^{(1)} ..... \theta_0 x_0^{(m)} + \theta_1 x_1^{(m)} + \theta_2 x_2^{(m)} + .... + \theta_n x_n^{(m)}]$$

But again using the definition of matrix multiplication, we can represent this more concisely:

$$h_\theta(x) = [\ \theta_0\ \theta_1 ... \theta_n\ ] \begin{bmatrix} x_0^{(1)} & x_0^{(2)} & \cdots & x_0^{(m)} \\ x_1^{(1)} & x_1^{(2)} & \cdots & x_1^{(m)} \\ \vdots & \vdots & & \vdots \\ x_n^{(1)} & x_n^{(2)} & \cdots & x_n^{(m)} \end{bmatrix} = \theta^T X$$

## Cost function

For the parameter vector θ (of type $R^{n+1}$ or in $R^{(n+1)\times 1}$), the cost function is:

$$J(\theta) = \frac{1}{2m} \sum_{i=1}^{m} \left( h_\theta(x^{(i)}) - y^{(i)} \right)^2$$

The vectorized version is:

$$J(\theta) = \frac{1}{2m} (X\theta - \vec{y})^T (X\theta - \vec{y}) \text{ , where } \vec{y} \text{ denotes the vector of all y values.}$$

## Gradient Descent for Multiple Variables

The gradient descent equation itself is generally the same form; we just have to repeat it for our 'n' features:

repeat until convergence: {

$$\theta_0 := \theta_0 - \alpha \frac{1}{m} \sum_{i=1}^{m} \left( (h_\theta(x^{(i)}) - y^{(i)}).x_0^{(i)} \right)$$

$$\theta_1 := \theta_1 - \alpha \frac{1}{m} \sum_{i=1}^{m} \left( (h_\theta(x^{(i)}) - y^{(i)}).x_1^{(i)} \right)$$

$$\theta_2 := \theta_2 - \alpha \frac{1}{m} \sum_{i=1}^{m} \left( (h_\theta(x^{(i)}) - y^{(i)}).x_2^{(i)} \right)$$

...... }

In other words:

repeat until convergence: {

$$\theta_j := \theta_j - \alpha \frac{1}{m} \sum_{i=1}^{m} \left( (h_\theta(x^{(i)}) - y^{(i)}).x_j^{(i)} \right) \text{ for j := 0,1,..,n } \}$$

The matrix notation (vectorized) of the Gradient Descent rule is:

$$\theta := \theta - \frac{\alpha}{m} X^T (X\theta - \vec{y})$$

<u>Normal Equation</u>

The "normal equation" is a version of finding the optimum without iteration.

$\theta=(X^TX)^{-1}X^Ty$

There is no need to do feature scaling with the normal equation.

The following is a comparison of gradient descent and the normal equation:

| Gradient Descent | Normal Equation |
|---|---|
| Need to choose alpha | No need to choose alpha |
| Needs many iterations | No need to iterate |
| Works well when n is large | Slow if n is very large |

**Table 3.1** – Comparison between Gradient Descent and Normal Equation

With the normal equation, computing the inversion has complexity $O(n^3)$. So if we have a very large number of features, the normal equation will be slow. According to Andrew Ng (Professor at Stanford) when n exceed 10,000 it might be a good time to go from a normal solution to an iterative process.

<u>Normal Equation Non-invertibility</u>

When implementing the normal equation in octave we want to use the 'pinv' function rather than 'inv.', i.e. we should use the pseudo-inverse rather than actual inverse.

$X^TX$ may be non-invertible. The common causes are:

- Redundant features, where two features are very closely related (i.e. they are linearly dependent)
- Too many features (e.g. $m \leq n$). In this case, delete some features or use "regularization" (explained later).

Solutions to the above problems include deleting a feature that is linearly dependent with another or deleting one or more features when there are too many features.

### 3.4  Logistic Regression

Now we are switching from regression problems to classification problems. We should not be confused by the named "Logistic Regression"; it is named that way for historical reasons and is actually an approach to classification problems, not regression problems.

<u>Classification</u>

Instead of our output vector y being a continuous range of values, it will only be 0 or 1 i.e. $y \in \{0,1\}$

0 is usually taken as "negative class" and 1 as "positive class", but you are free to assign any representation to it. We're only doing two classes for now, and it is called a "Binary Classification Problem."

One method is to use linear regression and map all predictions greater than 0.5 as a 1 and all less than 0.5 as a 0. This method doesn't work well because classification is not actually a linear function.

Hypothesis Representation

Our hypothesis should satisfy: $0 \leq h_\theta(x) \leq 1$

Our new form uses the "Sigmoid Function," also called the "Logistic Function", which is as follows:

$h_\theta(x) = g(\theta^T x) = g(z) = \frac{1}{1+e^{-z}}$

It is the same as the old hypothesis function (for linear regression), except that we are wrapping it in a call to g(), which is the Logistic Function.

$h_\theta$ will give us the probability that our output is 1. For example, $h_\theta(x) = 0.7$ gives us the probability of 70% that our output is 1.

$h_\theta(x) = P(y=1|x;\theta) = 1-P(y=0|x;\theta)$

Our probability that our prediction is 0 is just the opposite of our probability that it is 1 (e.g. if probability that it is 1 is 70%, then the probability that it is 0 is 30%).

Decision Boundary

In order to get our discrete 0 or 1 classification, we can translate the output of the hypothesis function as follows:

$h_\theta(x) \geq 0.5 \rightarrow y = 1$ ; $h_\theta(x) < 0.5 \rightarrow y = 0$

The way our logistic function g behaves is that when its input is greater than or equal to zero, its output is greater than or equal to 0.5 i.e. $g(z) \geq 0.5$ when $z \geq 0$. When its input is less than zero, its output is less than 0.5

Remember that:

$z=0$, $e^0=1$, $g(z)=1/2$

$z=\infty$, $e^{-\infty}=0$, $g(z)=1$

$z=-\infty$, $e^{\infty}=\infty$, $g(z)=0$

So if our input to g is $\theta^T X$, then that means:   $h_\theta(x) = g(\theta^T x) \geq 0.5$ when $\theta^T x \geq 0$

From these statements we can now say:  $\theta^T x \geq 0 \rightarrow y = 1$

$$\theta^T x < 0 \rightarrow y=0$$

Example:

$\theta = \begin{bmatrix} 5 \\ -1 \\ 0 \end{bmatrix}$,    $y = 1$  if $5 + (-1) x_1 + (0) x_2 \geq 0$   i.e.   if  $x_1 < 5$

The decision boundary is the line that separates the area where y=0 and where y=1. It is created by our hypothesis function.

Again, our hypothesis function need not be linear, and could be a function that describes a circle or any shape to fit our data.

Cost Function

We cannot use the same cost function that we use for linear regression because the Logistic Function will cause the output to be wavy, causing many local optima. In other words, it will not be a convex function.

Instead, our cost function for logistic regression looks like:

$$J(\theta) = \frac{1}{m} \sum_{i=1}^{m} \text{Cost}\big(h_{\theta}\big(x^{(i)}\big), y^{(i)}\big)$$

Cost($h_{\theta}$(x),y) = −log($h_{\theta}$(x))  if y = 1

Cost($h_{\theta}$(x),y) = −log(1−$h_{\theta}$(x))  if y = 0

The more our hypothesis is off from y, the larger the cost function's output. If our hypothesis is equal to y, then our cost is 0.

Cost($h_{\theta}$(x),y) = 0  if  $h_{\theta}$(x) = y

Cost($h_{\theta}$(x),y) $\rightarrow \infty$  if  y = 0 and $h_{\theta}$(x) $\rightarrow$ 1

Cost($h_{\theta}$(x),y) $\rightarrow \infty$  if  y = 1 and $h_{\theta}$(x) $\rightarrow$ 0

If our correct answer 'y' is 0, then the cost function will be 0 if our hypothesis function also outputs 0. If our hypothesis approaches 1, then the cost function will approach infinity.

If our correct answer 'y' is 1, then the cost function will be 0 if our hypothesis function outputs 1. If our hypothesis approaches 0, then the cost function will approach infinity.

<u>Simplified Cost Function and Gradient Descent</u>

We can compress our cost function's two conditional cases into one case:

Cost($h_{\theta}$(x),y) = − y log($h_{\theta}$(x)) − (1−y) log(1−$h_{\theta}$(x))

Notice that when y is equal to 1, then the second term ((1−y) log(1−$h_{\theta}$(x))) will be negated and will not affect the result. If y is equal to 0, then the first term (− y log($h_{\theta}$(x))) will be negated and will not affect the result.

We can fully write out our entire cost function as follows:

$$J(\theta) = - \frac{1}{m} \sum_{i=1}^{m} \left[ y^{(i)} \log \left( h_\theta\big(x^{(i)}\big) \right) + \left( 1 - y^{(i)} \right) \log \left( 1 - h_\theta\big(x^{(i)}\big) \right) \right]$$

A vectorized implementation is:

$$J(\theta) = - \frac{1}{m} \left[ \log(g(X\theta))^\top \, y + \log(1-g(X\theta))^\top \, (1-y) \right]$$

Gradient Descent

Remember that the general form of gradient descent is:

Repeat until convergence: {

$$\theta_j := \theta_j - \alpha \frac{\partial}{\partial \theta_j} J(\theta) \quad \}$$

We can work out the derivative part using calculus to get:

Repeat until convergence: { $\theta_j := \theta_j - \frac{\alpha}{m} \sum_{i=1}^{m} \left( \big(h_\theta\big(x^{(i)}\big) - y^{(i)}\big) . x_j^{(i)} \right)$ }

Notice that this algorithm is identical to the one we used in linear regression, but the hypothesis function is different for linear and logistic regression. We still have to simultaneously update all values in theta.

A vectorized implementation is:

$$\theta := \theta - \frac{\alpha}{m} X^\top \left( g(X\theta) - \vec{y} \right)$$

## 3.5  Regularization

The Problem of Overfitting

Regularization is designed to address the problem of overfitting.

High bias or underfitting is when the form of our hypothesis maps poorly to the trend of the data. It is usually caused by a function that is too simple or uses too few features.

At the other extreme, overfitting or high variance is caused by a hypothesis function that fits the available data but does not generalize well to predict new data. It is usually caused by a complicated function that creates a lot of unnecessary curves and angles unrelated to the data.

This terminology is applied to both linear and logistic regression.

There are two main options to address the issue of overfitting:

1. Reduce the number of features.

   o Manually select which features to keep.

   o Use a model selection algorithm.

2. Regularization

   o Keep all the features, but reduce the parameters $\theta_j$

Regularization works well when we have a lot of slightly useful features.

Regularized Linear Regression

Gradient Descent

We will modify our gradient descent function to separate out $\theta_0$ from the rest of the parameters because we do not want to penalize $\theta_0$.

Repeat until convergence: {

$$\theta_0 := \theta_0 - \frac{\alpha}{m} \sum_{i=1}^{m} \left( \left( h_\theta\left(x^{(i)}\right) - y^{(i)}\right).x_0^{(i)} \right)$$

$$\theta_j := \theta_j - \alpha \left[ \frac{1}{m} \sum_{i=1}^{m} \left( \left( h_\theta(x^{(i)}) - y^{(i)} \right). x_j^{(i)} \right) + \frac{\lambda}{m} \theta_j \right] \quad j \in \{1,2...n\}$$

}

The term $\frac{\lambda}{m} \theta_j$ performs our regularization.

With some manipulation our update rule can also be represented as:

$$\theta_j := \theta_j \left( 1 - \alpha \frac{\lambda}{m} \right) - \frac{\alpha}{m} \sum_{i=1}^{m} \left( \left( h_\theta(x^{(i)}) - y^{(i)} \right). x_j^{(i)} \right)$$

The first term in the above equation, $\left( 1 - \alpha \frac{\lambda}{m} \right)$ will always be less than 1. Intuitively you can see it as reducing the value of $\theta_j$ by some amount on every update.

Notice that the second term is now exactly the same as it was before.

Regularized Logistic Regression

We can regularize logistic regression in a similar way that we regularize linear regression. Let's start with the cost function.

Cost Function

Recall that our cost function for logistic regression was:

$$J(\theta) = - \frac{1}{m} \sum_{i=1}^{m} \left[ y^{(i)} \log \left( h_\theta(x^{(i)}) \right) + \left( 1 - y^{(i)} \right) \log \left( 1 - h_\theta(x^{(i)}) \right) \right]$$

We can regularize this equation by adding a term to the end:

$$J(\theta) = - \frac{1}{m} \sum_{i=1}^{m} \left[ y^{(i)} \log \left( h_\theta(x^{(i)}) \right) + \left( 1 - y^{(i)} \right) \log \left( 1 - h_\theta(x^{(i)}) \right) \right] + \frac{\lambda}{2m} \sum_{j=1}^{n} \theta_j^2$$

Note Well: The second sum $\sum_{j=1}^{n} \theta_j^2$ means to explicitly exclude the bias term $\theta_0$

<u>Gradient Descent</u>

Just like with linear regression, we will want to separately update $\theta_0$ and the rest of the parameters because we do not want to regularize $\theta_0$.

Repeat until convergence: $\{$ $\theta_0 := \theta_0 - \frac{\alpha}{m} \sum_{i=1}^{m} \big( (h_\theta(x^{(i)}) - y^{(i)}).x_0^{(i)} \big)$

$\theta_j := \theta_j - \alpha \left[ \frac{1}{m} \sum_{i=1}^{m} \big( (h_\theta(x^{(i)}) - y^{(i)}).x_j^{(i)} \big) + \frac{\lambda}{m} \theta_j \right],$ j=1,2...n

$\}$

This is identical to the gradient descent function presented for linear regression.

## 3.6  Support Vector Machines

<u>Optimization Objective</u>

The Support Vector Machine (SVM) is yet another type of supervised machine learning algorithm. It is sometimes cleaner and more powerful.

Recall that in logistic regression, we use the following rules:

if y=1, then $h_\theta(x) \approx 1$ and $\theta^T x > 0$

if y=0, then $h_\theta(x) \approx 0$ and $\theta^T x < 0$

Recall the cost function for (unregularized) logistic regression:

$J(\theta) = -\frac{1}{m} \sum_{i=1}^{m} \left[ y^{(i)} \log \big( h_\theta(x^{(i)}) \big) + \big( 1 - y^{(i)} \big) \log \big( 1 - h_\theta(x^{(i)}) \big) \right]$

$\quad = -\frac{1}{m} \sum_{i=1}^{m} \left[ y^{(i)} \log \big( \frac{1}{1 + e^{-\theta^T x^{(i)}}} \big) + \big( 1 - y^{(i)} \big) \log \big( 1 - \frac{1}{1 + e^{-\theta^T x^{(i)}}} \big) \right]$

To make a support vector machine, we will modify the first term of the cost function [ $h_\theta(x^{(i)}) = \frac{1}{1 + e^{-\theta^T x^{(i)}}}$ ] so that when $\theta^T x$ (from now on, we shall refer to this as z) is greater than 1, it outputs 0. Furthermore, for values of z less than 1, we shall use a straight decreasing line instead of the sigmoid curve.(In the literature, this is called a hinge loss function.)

Similarly, we modify the second term of the cost function [ $1 - h_\theta(x^{(i)}) = 1 - \frac{1}{1 + e^{-\theta^T x^{(i)}}}$ ] so that when z is less than -1, it outputs 0. We also modify it so that for values of z greater than -1, we use a straight increasing line instead of the sigmoid curve.

We shall denote these as $cost_1(z)$ and $cost_0(z)$ respectively (note that $cost_1(z)$ is the cost for classifying when y=1, and $cost_0(z)$ is the cost for classifying when y=0), and we may define them as follows (where k is an arbitrary constant defining the magnitude of the slope of the line):

$z = \theta^T x$ , $cost_0(z) = \max(0, k(1+z))$ , $cost_1(z) = \max(0, k(1-z))$

Recall the full cost function from (regularized) logistic regression:

$J(\theta) = \frac{1}{m}\sum_{i=1}^{m}\left[y^{(i)}\left(-\log\left(h_\theta(x^{(i)})\right)\right) + (1 - y^{(i)})\left(-\log\left(1 - h_\theta(x^{(i)})\right)\right)\right] + \frac{\lambda}{2m}\sum_{j=1}^{n}\theta_j^2$

Note that the negative sign has been distributed into the sum in the above equation.

We may transform this into the cost function for support vector machines by substituting $cost_0(z)$ and $cost_1(z)$:

$J(\theta) = \frac{1}{m}\sum_{i=1}^{m}\left[y^{(i)}cost_1\left(\theta^T x^{(i)}\right) + (1 - y^{(i)})cost_0\left(\theta^T x^{(i)}\right)\right] + \frac{\lambda}{2m}\sum_{j=1}^{n}\theta_j^2$

We can optimize this a bit by multiplying this by m (thus removing the m factor in the denominators). Note that this does not affect our optimization, since we're simply multiplying our cost function by a constant (for example, minimizing $(u-5)2+1$ gives us 5; multiplying it by 10 to make it $10(u-5)2+10$ still gives us 5 when minimized).

$$J(\theta) = \sum_{i=1}^{m} \left[ y^{(i)}\text{cost}_1\left(\theta^T x^{(i)}\right) + \left(1 - y^{(i)}\right)\text{cost}_0\left(\theta^T x^{(i)}\right) \right] + \frac{\lambda}{2}\sum_{j=1}^{n} \theta_j^2$$

Furthermore, convention dictates that we regularize using a factor C, instead of $\lambda$, as given by the following:

$$J(\theta) = C\sum_{i=1}^{m} \left[ y^{(i)}\text{cost}_1\left(\theta^T x^{(i)}\right) + \left(1 - y^{(i)}\right)\text{cost}_0\left(\theta^T x^{(i)}\right) \right] + \frac{1}{2}\sum_{j=1}^{n} \theta_j^2$$

This is equivalent to multiplying the equation by $C = \frac{1}{\lambda}$, and thus results in the same values when optimized. Now, when we wish to regularize more, we decrease C, and when we wish to regularize less, we increase C.

Finally, note that the hypothesis of the Support Vector Machine is not interpreted as the probability of y being 1 or 0 (as it is for the hypothesis of logistic regression). Instead, it outputs either 1 or 0. (In technical terms, it is a discriminant function)

$$h_\theta(x) = \begin{cases} 1 & \text{if } \theta^T x \geq 0 \\ 0 & \text{otherwise} \end{cases}$$

<u>Large Margin Intuition</u>

A useful way to think about Support Vector Machines is to think of them as Large Margin Classifiers.

If y=1, we want $\theta^T x \geq 1$ (not just $\geq 0$)

If y=0, we want $\theta^T x \leq -1$ (not just $< 0$)

Now when we set our constant C to a very large value (e.g. 100,000), our optimizing function will constrain $\theta$ and the equation involving the sum of the cost of each example equals 0.

We impose the following constraints on $\theta$:

$\theta^T x \geq 1$, if y=1 and $\theta^T x \leq -1$, if y=0.

If C is very large, then we must choose $\theta$ parameters such that:

$\sum_{i=1}^{m} \left[ y^{(i)} \text{cost}_1\left(\theta^T x^{(i)}\right) + \left(1 - y^{(i)}\right)\text{cost}_0\left(\theta^T x^{(i)}\right)\right] = 0$

This reduces our cost function to:

$J(\theta) = C \cdot 0 + \frac{1}{2}\sum_{j=1}^{n} \theta_j^2 = \frac{1}{2}\sum_{j=1}^{n} \theta_j^2$

Recall the decision boundary from logistic regression (the line separating the positive and negative examples). In SVMs, the decision boundary has the special property that it is as far away as possible from both the positive and the negative examples.

The distance of the decision boundary to the nearest example is called the margin. Since SVMs maximize this margin, it is often called a Large Margin Classifier.

The SVM will separate the negative and positive examples by a large margin.

This large margin is only achieved when C is very large.

Data is linearly separable when a straight line can separate the positive and negative examples.

If we have outlier examples that we don't want to affect the decision boundary, then we can reduce C.

Increasing and decreasing C is similar to increasing and decreasing λ because it can simplify our decision boundary.

## 3.7  Prediction Performance

After the construction of mathematical models with multi-regression and support vector regression, the models are used to predict the true label value. During the prediction process, for any feature group, the entire data set is split at random into 80% training set and 20% test set. These feature vectors from the training set are used to construct the mathematical models by using Multi-Regression or Support Vector Regression with linear kernel to predict failure load values. Once the mathematical models are constructed, the feature vectors from the test set are fetched into the models to predict the failure load values, and these predicted failure load values are compared with the ground truth by Root-Mean Square Error (RMSE), defined as:

$$\text{RMSE} = \sqrt{(\text{FL}_{\text{pred}} - \text{FL}_{\text{true}})^2}$$ , where $\text{FL}_{\text{pred}}$ is the predicted failure load and $\text{FL}_{\text{true}}$ is the true failure load for the test set.

Fifty iterations of this prediction process are performed and the RMSE measured from different feature groups such as Isotropic Minkowski Functionals (IMF) and Anisotropic Minkowski Functionals (AMF) using different regression methods (Multi-Regression or Support Vector Regression with linear kernel) is compared to the RMSE with the standard approach, which uses BMD mean with a Multi-Regression model. A Wilcoxon signed-rank test was used to compare two RMSE distributions corresponding to the prediction performance of different features. Significance thresholds were adjusted for multiple comparisons using the Holm-Bonferroni correction to achieve an overall type I error rate (significance level) less than $\alpha$ (where $\alpha = 0.05$) [47, 48].

# Chapter 4

## Bone Strength Prediction: Performance Results

The prediction performance of different texture analysis and machine learning techniques discussed previously are compared with the current clinical standard.

### 4.1  Identification of Femur Region for Analysis

The fundamental BMD statistics distributions of the dataset were examined (Table 4.1) to investigate the correlation between BMD measurements from different regions and FL. In addition, the FL was estimated using multi-regression analysis (Fig 4.1) to identify the ideal candidate for further analysis.

|  | Max | Min | Mean | SD | r with FL |
|---|---|---|---|---|---|
| Age (years) | 100 | 52 | 79.39 | 10.57 | - |
| Failure Load (kN) | 8.156 | 0.664 | 3.943 | 1.557 | - |
| BMD.mean Head (mg/cm$^3$) | 406.91 | 57.33 | 218.96 | 64.73 | 0.706 |
| BMD.mean Neck (mg/cm$^3$) | 225 | -46.22 | 44.98 | 53.38 | 0.467 |
| BMD.mean Troch (mg/cm$^3$) | 226 | -35.52 | 70.79 | 52.94 | 0.596 |

**Table 4.1:** Values of investigated parameters for femur specimens. Representative statistical values of Age, Failure Load (FL) and the mean BMD of three femur specimen regions are listed. The correlation between the mean BMD from different regions with FL are calculated. BMD mean of head region has higher correlation. Adapted from [44]

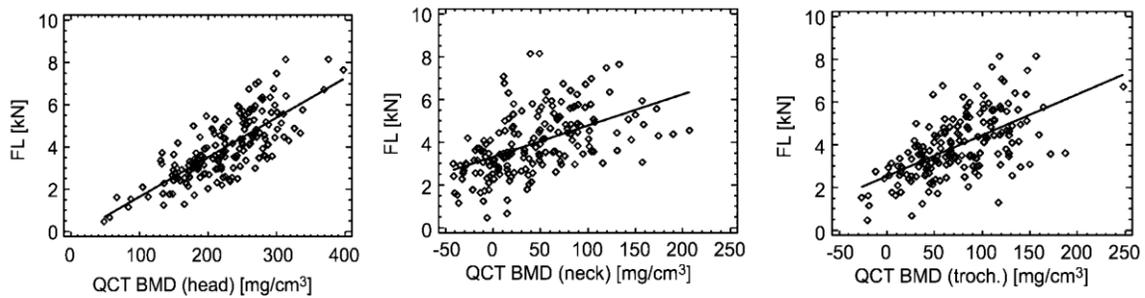

**Figure 4.1:** Scatterplots show relationships between failure load *(FL)* and BMD. Coefficients *(r)* for correlations with failure load were as follows: (a) 0.706 for correlation with quantitative CT *(QCT)* BMD in femur head, (b) 0.467 for correlation with quantitative CT BMD in femur neck, and (c) 0.596 for correlation with quantitative CT BMD in trochanter *(troch.).* All correlations were significant (p < 0.001). Each solid line represents the fit to a linear regression model. Adapted from [43].

Since the head region BMD yields the highest correlation (r = 0.706), it was selected for further texture feature analysis and prediction performance tests. So in our future analysis, we will only be using the head region.

## 4.2   Conventional Statistical Features

The DXA BMD value was extracted from the trochanter, neck, ward and shaft regions of the DXA image of each proximal femur specimen, and the mean of these 4 BMD values corresponding to these 4 regions, denoted as total DXA BMD or simply DXA BMD, was used as a feature to construct mathematical models for prediction of biomechanical strength (failure load).

Using DXA BMD and Multi-Regression, the prediction performance results obtained were **RMSE = 0.960 ± 0.131**

Using DXA BMD and Support Vector Regression (SVR) with linear kernel, the prediction performance results obtained were **RMSE = 0.959 ± 0.132**

Thus, we can conclude that for DXA BMD feature, both Multi-Regression and Support Vector Regression (SVR) gives equally good prediction performance.

## 4.3  Isotropic Minkowski Functionals

In order to obtain the Isotropic Minkowski Functionals (IMFs), we first need to threshold the BMD image to convert it into a black and white image. In our case, we have empirically chosen the threshold BMD value to be 400.

Then we need to identify optimal values for free parameters (in this case the kernel size) in order to obtain the best prediction performance. We have used a number of different kernel sizes (ranging from 5x5x5 to 19x19x19 in increments of 2) and fixed histogram bin size of 10 in order to get the IMF features. We have then evaluated the prediction performance for each kernel size and each IMF feature i.e. Volume, Surface, Mean Breadth and Euler Characteristic.

The following table lists the prediction performance obtained using the Isotropic Minkowski Functionals (IMF) and DXA BMD:-

| Feature Groups | Multi-Regression (RMSE) | SVR (RMSE) |
|---|---|---|
| DXA BMD | **0.960 ± 0.131** | 0.959 ± 0.132 |
| IMF.volume | 1.612 ± 0.163 | 1.585 ± 0.167 |
| DXA BMD + IMF.volume | 0.999 ± 0.113 | 0.992 ± 0.140 |
| IMF.surface | 1.701 ± 0.249 | 1.631 ± 0.200 |
| DXA BMD + IMF.surface | 1.003 ± 0.122 | 0.995 ± 0.146 |
| IMF.mean_breadth | 1.695 ± 0.226 | 1.625 ± 0.190 |
| DXA BMD + IMF.mean_breadth | 1.017 ± 0.125 | 0.985 ± 0.132 |
| IMF.euler | 1.669 ± 0.208 | 1.600 ± 0.183 |
| DXA BMD + IMF.euler | 1.026 ± 0.133 | **0.981 ± 0.134** |

**Table 4.2:** Table showing the prediction performance (RMSE) of Feature Groups DXA BMD and Isotropic Minkowski Functionals used in conjunction with Multi-Regression and Support Vector Regression with linear kernel.

From the above table, we can see that Isotropic Minkowski Functionals itself is not very good. The best prediction performance for Isotropic Minkowski Functionals alone is given by IMF.volume with RMSE = 1.585 ± 0.167 which is significantly lower ($p < 0.05$) than the standard approach of using DXA BMD and multi-regression (RMSE = 0.960 ± 0.131).

## 4.4 Anisotropic Minkowski Functionals

In order to extract the Anisotropic Minkowski Functionals, we have to first threshold the BMD image to obtain a black and white image, similarly as with Isotropic Minkowski Functionals. As before, we have empirically chosen our threshold BMD value to be 400.

Then we have to optimize our free parameters (in this case it is the kernel size and the ratio between the standard deviation of the Gaussian kernel in the principal direction and its two orthogonal directions) to obtain the best prediction performance. For this reason, we have chosen a number of different kernel sizes ranging from 5x5x5 to 19x19x19 in increments of 2, and the ratio between the standard deviations have been chosen as 2, 4 and 8.

The following table lists the prediction performance obtained using the Anisotropic Minkowski Functionals (AMF) and DXA BMD :

| Feature Groups | Multi-Regression (RMSE) | SVR (RMSE) |
|---|---|---|
| DXA BMD | **0.960 ± 0.131** | 0.959 ± 0.132 |
| AMF.volume | 1.060 ± 0.126 | 1.007 ± 0.105 |
| DXA BMD + AMF.volume | 0.909 ± 0.111 | 0.880 ± 0.112 |
| AMF.surface | 1.051 ± 0.130 | 1.018 ± 0.120 |
| DXA BMD + AMF.surface | 0.921 ± 0.116 | 0.894 ± 0.115 |
| AMF.mean_breadth | 1.056 ± 0.154 | 0.998 ± 0.115 |
| DXA BMD + AMF.mean_breadth | 0.995 ± 0.128 | 0.904 ± 0.101 |
| AMF.euler | 0.966 ± 0.128 | 0.904 ± 0.105 |
| DXA BMD + AMF.euler | 0.898 ± 0.116 | **0.838 ± 0.092** |

**Table 4.3:** Table showing the prediction performance (RMSE) of Feature Groups DXA BMD and Anisotropic Minkowski Functionals used in conjunction with Multi-Regression and Support Vector Regression with linear kernel.

The following 8 figures will show a comparison of the prediction performance (measured with RMSE) using features such as DXA BMD, Anisotropic Minkowski Functionals and Isotropic Minkowski Functionals.

[Note: In the following figures, the RMSE distribution obtained using Multi-Regression and Support Vector Regression with linear kernel is shown in red and blue colors respectively. For each RMSE distribution, the central mark corresponds to the median of the distribution and the top and bottom edges correspond to the 75th and 25th percentile respectively. The red horizontal line corresponds to the performance achieved with the standard approach (mean BMD with multi-regression). The blue line corresponds to the best performance achieved for the feature groups used in each figure.

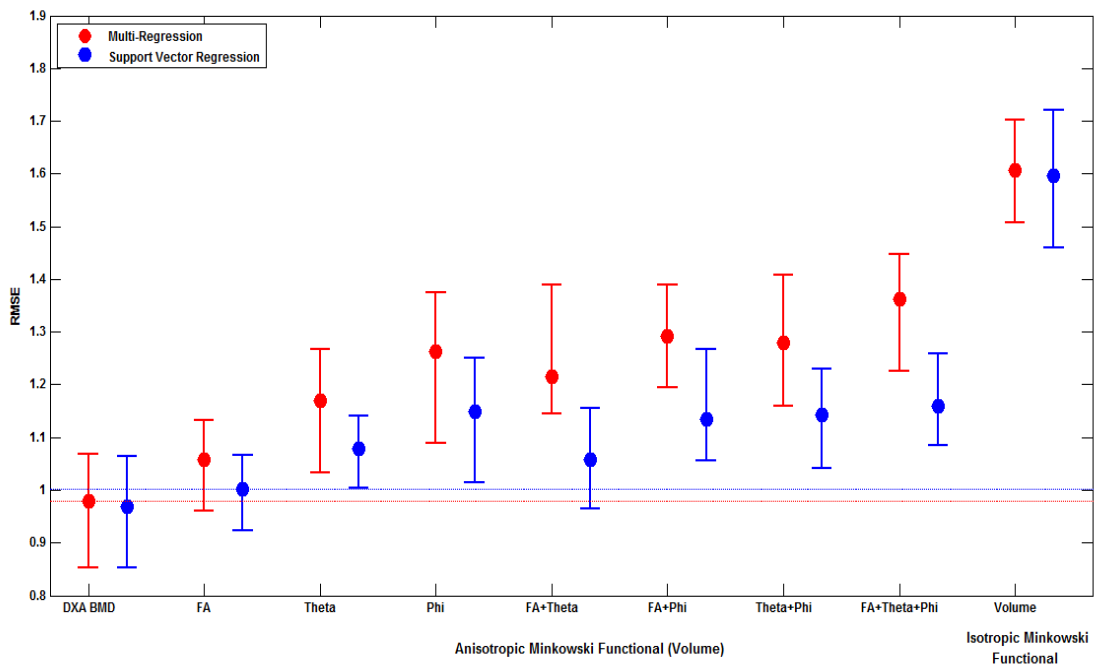

**Figure 4.2:** Figure showing the prediction performance (RMSE) of Feature Groups DXA BMD, AMF volume, IMF volume.

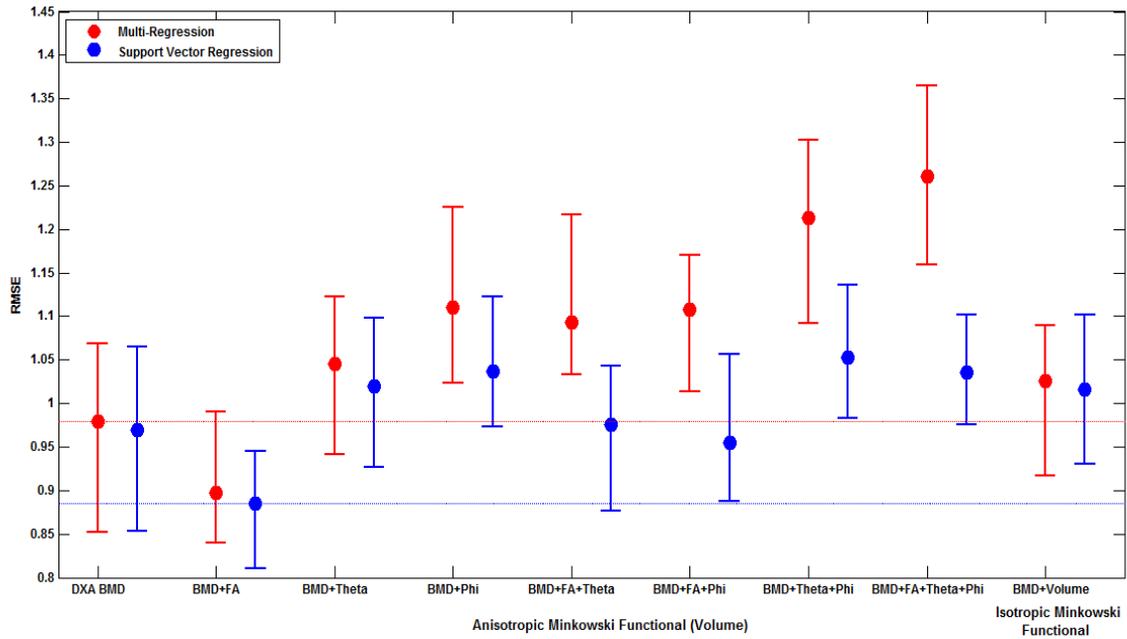

**Figure 4.3:** Figure showing the prediction performance (RMSE) of Feature Groups DXA BMD, DXA BMD + AMF volume, DXA BMD + IMF volume.

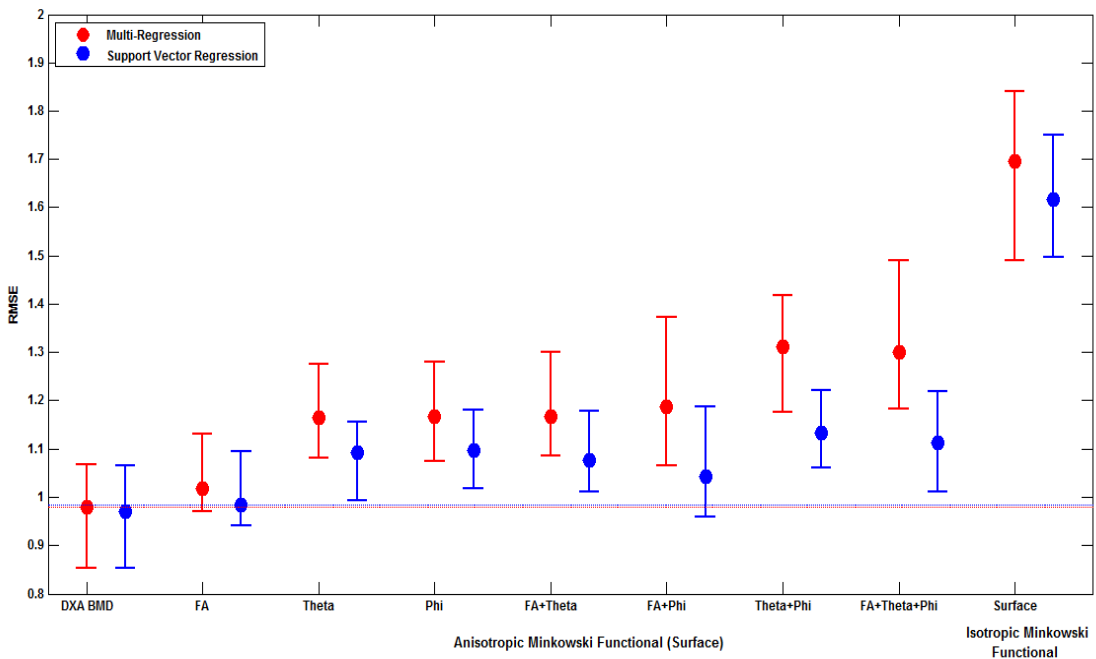

**Figure 4.4:** Figure showing the prediction performance (RMSE) of Feature Groups DXA BMD, AMF surface, IMF surface.

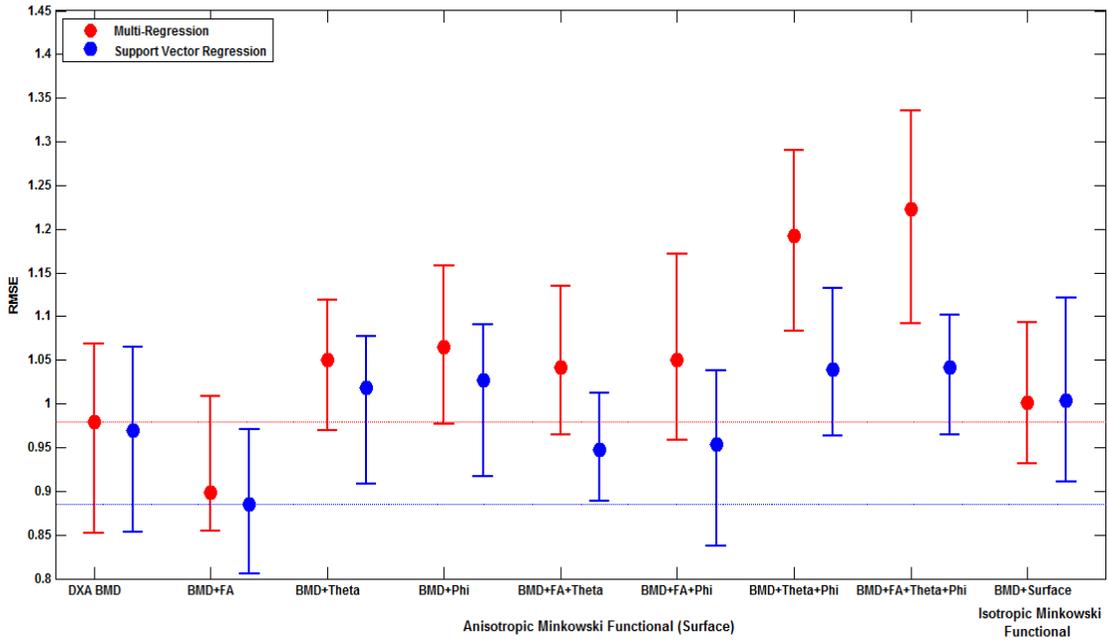

**Figure 4.5:** Figure showing the prediction performance (RMSE) of Feature Groups DXA BMD, DXA BMD + AMF surface, DXA BMD + IMF surface.

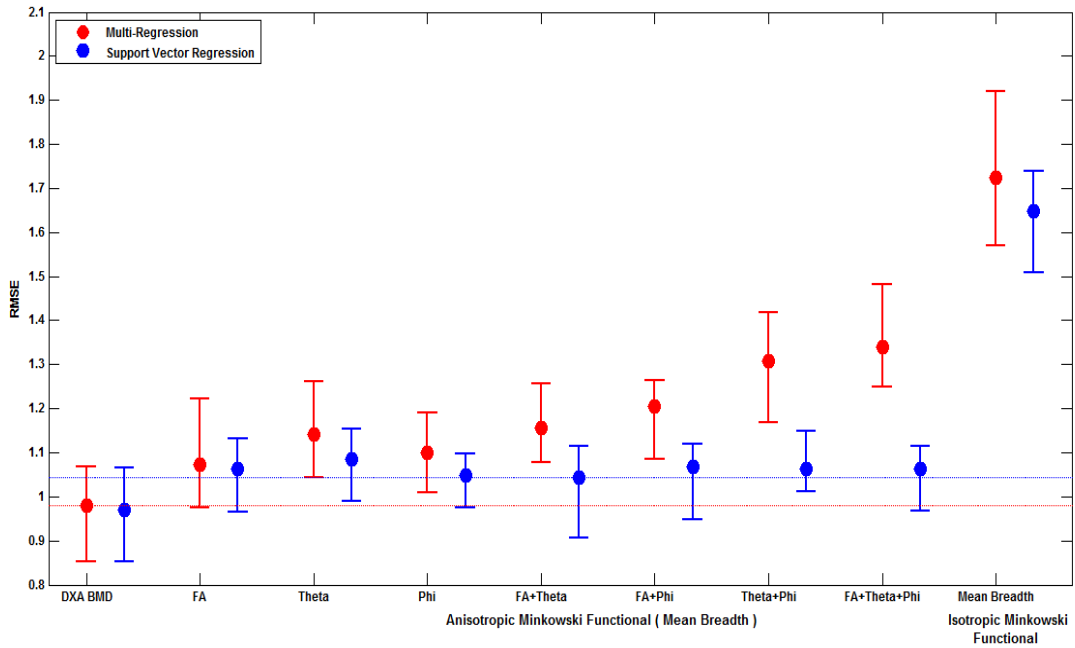

**Figure 4.6:** Figure showing the prediction performance (RMSE) of Feature Groups DXA BMD, AMF mean breadth, IMF mean breadth.

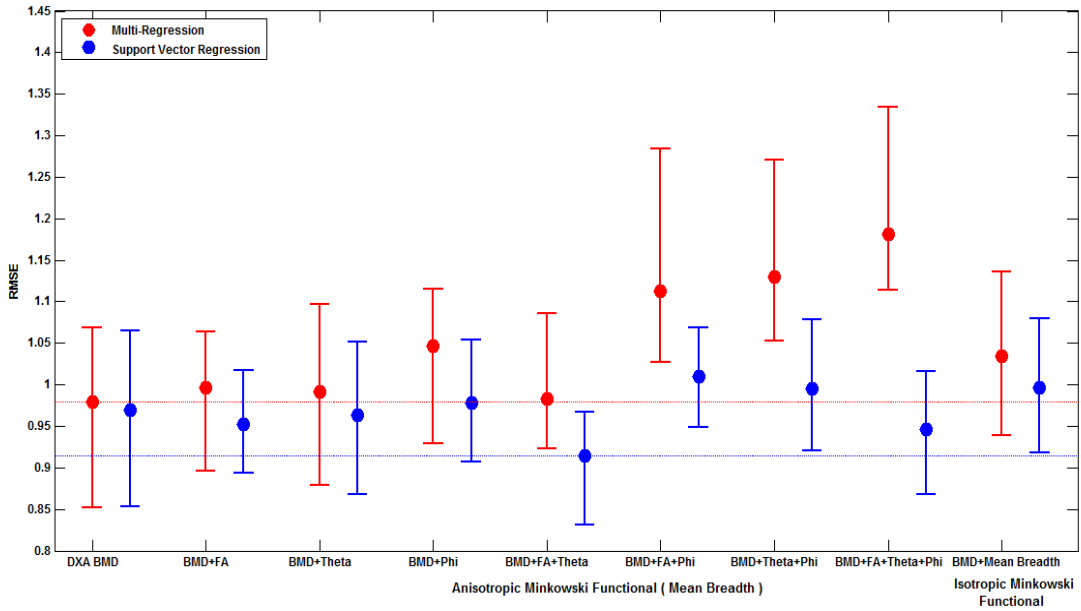

**Figure 4.7:** Figure showing the prediction performance (RMSE) of Feature Groups DXA BMD, DXA BMD+ AMF mean breadth, DXA BMD + IMF mean breadth.

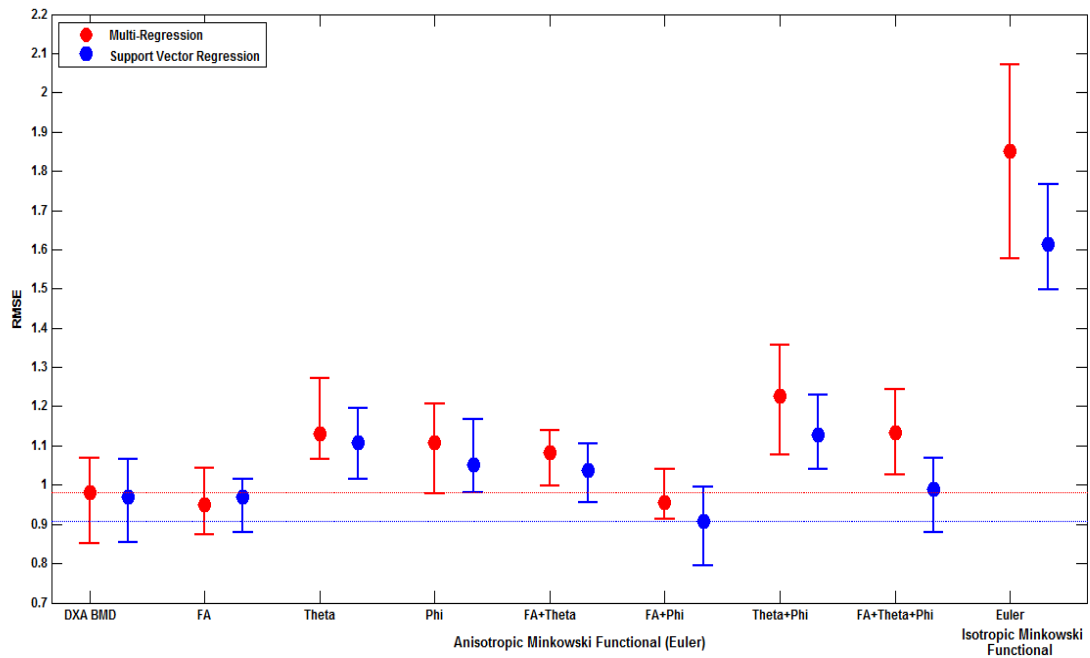

**Figure 4.8:** Figure showing the prediction performance (RMSE) of Feature Groups DXA BMD, AMF euler characteristic, IMF euler characteristic.

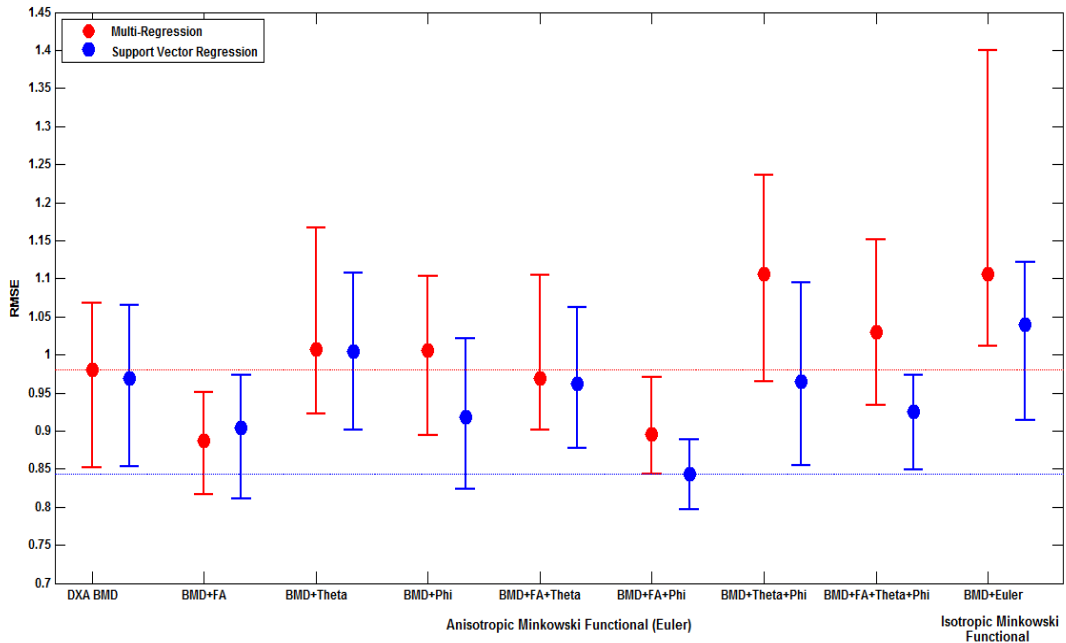

**Figure 4.9:** Figure showing the prediction performance (RMSE) of Feature Groups DXA BMD, DXA BMD + AMF euler characteristic, DXA BMD + IMF euler characteristic.

## 4.5 Prediction Performance Comparison and Conclusion

While looking at the above 8 figures, the general trend is that if we combine DXA BMD mean with any feature, be it IMF or AMF, the prediction performance of the combination is always better than that feature alone. Also, Support Vector Regression in general gives better prediction performance (in terms of lower RMSE) than Multi-Regression.

The best prediction performance using Anisotropic Minkowski Functionals (AMF) alone was obtained for a combination of FA and Phi feature vectors corresponding to Euler Characteristic (kernel size = 17, xyratio = 4). The best prediction performance obtained using Isotropic Minkowski Functionals (IMF) alone was obtained for Volume feature vector with kernel size 5.

The summary of the prediction performance obtained using different feature groups are shown in the following table:-

| Feature Groups | Multi-Regression (RMSE) | SVR (RMSE) |
|---|---|---|
| DXA BMD | **0.960 ± 0.131** | 0.959 ± 0.132 |
| Best IMF feature alone | 1.612 ± 0.163 | 1.585 ± 0.167 |
| Best AMF feature alone | 0.966 ± 0.116 | 0.904 ± 0.105 |
| Best combination of DXA BMD and IMF feature | 1.026 ± 0.133 | 0.981 ± 0.134 |
| Best combination of DXA BMD and AMF feature | 0.898 ± 0.116 | **0.838 ± 0.092** |

**Table 4.4:** Table showing the prediction performance (RMSE) of Feature Groups DXA BMD, best IMF and AMF alone, and best combination of DXA BMD and IMF and DXA BMD and AMF used in conjunction with Multi-Regression and Support Vector Regression with linear kernel.

The final conclusions I have drawn from all the results obtained are as follows:-

- Overall best prediction performance was obtained using the best combination of DXA BMD and AMF feature, and which was significantly better than DXA BMD alone ($p < 10^{-4}$)

- Prediction performance obtained using best AMF feature alone was significantly better than best IMF feature alone ($p < 10^{-4}$)

- Prediction performance obtained using best AMF feature alone was significantly better than DXA BMD alone ($p < 0.05$)

**Chapter 5**

**Discussion**

The correlation of QCT-derived mean BMD, a descriptor of bone mineral content in the trabecular bone of proximal femur specimens, to bone strength or failure load (FL) has been established [20,21]. However, one drawback with this measure is its inability to adequately characterize the micro-architecture of the femoral trabecular compartment. Previous studies have shown supplementing DXA BMD measurements (current clinical standard) with features that characterize the trabecular bone texture variation and micro-architecture can improve the corresponding correlation to bone strength on high-resolution MRI [28,29,37,38] and multi-detector CT [20-22,39]. We specifically investigate the ability of such features in predicting the failure load of the femur specimen through a computer-aided diagnosis approach involving regression analysis. Our results suggest that the inclusion of texture features derived from Anisotropic Minkowski Functionals in addition to DXA BMD significantly improves the accuracy of FL prediction in such proximal femur specimens. This suggests that such descriptors of trabecular bone quality and trabecular texture variation have significant potential to aid clinicians in predicting femoral fracture risk in patients suffering from osteoporosis.

As seen in Figure 4.1, the correlation between measured FL and QCT-derived mean BMD for head, neck and trochanter regions was the highest for the head region. Since these findings suggest that regional characterization of femoral trabeculae could play a significant role in FL prediction, subsequent feature extraction and regression has been focused only on the head region of the femur.

I have used Isotropic Minkowski Functionals (IMF) and Anisotropic Minkowski Functionals (AMF) as the texture features for characterizing trabecular bone micro-architecture in the femoral head region. Both IMF and AMF have the ability to characterize the local structure within a region/volume of interest. But AMF captures some more information (i.e. anisotropy in patterns) about the local structure content in comparison to IMF.

Our results show that the prediction performance obtained using best AMF feature alone (RMSE = 0.904 ± 0.105) was significantly better ($p < 10^{-4}$) than using best IMF feature alone (1.585 ± 0.167). This is due to the fact that the anisotropy i.e. the directional patterns in bone structure (captured by AMF and not by IMF) is highly correlated to its bone strength (FL).

The overall best prediction performance was obtained using DXA BMD + AMF (RMSE = 0.838 ± 0.092), which was significantly better ($p < 10^{-4}$) than using DXA BMD alone (RMSE = 0.960 ± 0.131), and also significantly better ($p < 10^{-4}$) than using AMF alone (RMSE = 0.904 ± 0.105). We know that DXA BMD captures the bone mineral density and bone mineral content information from not only the trabecular bone but also from the cortical bone. So DXA BMD is able to gain insight about bone stability. Anisotropic Minkowski Functionals (AMF), on the other hand, are able to characterize the structure content (not the bone mineral content) of the trabecular bone. Therefore we find that AMF and DXA BMD capture complementary information about the femoral bone. If we combine AMF and DXA BMD together into a single feature vector, the amount of information about the femoral bone in their combination is much more than these features individually. Thus, it is quite obvious that if DXA BMD and AMF are combined together into a single feature vector for prediction performance, they will give the overall best prediction performance.